\documentclass[preprintnumbers,article,amsmath,amssymb,floatfix,10pt,prd,onecolumn,superscriptaddress,nofootinbib]{revtex4-2}
\usepackage{bm}
\usepackage{amsfonts}
\usepackage{latexsym}
\usepackage[latin1]{inputenc}
\usepackage{graphicx}
\usepackage{amsmath}
\usepackage{palatino}
\usepackage{mathpazo}
\usepackage{textcomp}
\usepackage{textcomp}
\usepackage{mathrsfs}
\usepackage{float}
\usepackage{booktabs}
\usepackage{dcolumn}
\usepackage{ragged2e}
\usepackage{hyperref}
\usepackage{tensor}
\usepackage[version=3]{mhchem}
\hypersetup{colorlinks,citecolor=blue}
\hypersetup{colorlinks=true,linkcolor=red,filecolor=magenta,    urlcolor=blue}
\usepackage{amsmath}
\usepackage{xcolor}
\usepackage{enumitem}
\usepackage{orcidlink}
\usepackage{epsfig}
\usepackage{subfigure}
\usepackage{commath}
\usepackage{caption}

\begin{document}

\title{Exploring Late-Time Cosmic Acceleration in VCDM Cosmology}
\author{Sai Swagat Mishra\orcidlink{0000-0003-0580-0798}}
\email{saiswagat009@gmail.com}
\affiliation{Department of Mathematics, School of Computer Science and Artificial Intelligence, SR University, Warangal 506371, Telangana, India}%

\author{Soumya Kanta Bhoi\orcidlink{0009-0000-4392-317X}}
\email{soumyakanta.1711@gmail.com}
\affiliation{Department of Mathematics, Birla Institute of Technology and Science, Pilani, Hyderabad Campus, Jawahar Nagar, Kapra Mandal, Medchal District, Telangana 500078, India}%

\author{P.K. Sahoo\orcidlink{0000-0003-2130-8832}}
\email{pksahoo@hyderabad.bits-pilani.ac.in}
\affiliation{Department of Mathematics, Birla Institute of Technology and Science, Pilani, Hyderabad Campus, Jawahar Nagar, Kapra Mandal, Medchal District, Telangana 500078, India}
\begin{abstract}
    In this work, we have considered a minimally modified gravity theory that effectively reproduces VCDM-like behavior to investigate its cosmological implications. The model parameters are constrained using a combination of CC, RSD, DESI BAO DR2, and Union3 datasets. The model parameters are constrained using an MCMC framework, ensuring a robust estimation of credible intervals. We examine both background and perturbation-level observables, analyzing the Hubble parameter, deceleration parameter, effective equation of state, distance modulus, and the growth rate of cosmic structures through the \(f\sigma_8(z)\) observable.
    Our results show that the model successfully reproduces the observed expansion history, featuring a smooth transition in the Hubble evolution around \(z \simeq 0.3\), and consistent behaviour of cosmological parameters. The model outperforms \(\Lambda\)CDM in the statistical comparisons for the full dataset combination (CC+RSD+DESI DR2+Union3). These results highlight the potential of minimally modified gravity theories with VCDM-like dynamics as consistent and competitive alternatives to the standard cosmological paradigm.
    
\end{abstract}
\maketitle

\section{Introduction}
In 1998, observations of distant Type Ia supernovae by Riess et al.\cite{SupernovaSearchTeam:1998fmf} and Perlmutter et al.\cite{SupernovaCosmologyProject:1998vns} revealed that the expansion of the Universe is accelerating. This unexpected discovery posed a serious challenge to the prevailing cosmological models of the time, such as the Einstein-de Sitter, Open CDM, Closed CDM, and Standard CDM models, which only incorporated matter and radiation. These models did not account for any component that could drive accelerated expansion, implying their incompleteness.

To resolve this, cosmologists introduced a new component: the cosmological constant $\Lambda$, interpreted as Dark Energy. This idea was motivated by the presence of a constant term in Einstein's field equations for a static universe and further supported by the inflationary paradigm \cite{Guth:1980zm}, which demonstrated that a scalar field vacuum energy, also called the ``false vacuum'', could induce rapid expansion in the early universe. It was thus proposed that vacuum energy could produce gravitational effects similar to a cosmological constant. The resulting framework, known as the $\Lambda$CDM model, includes $\Lambda$ to explain the observed late-time acceleration and Cold Dark Matter (CDM) to account for large-scale structure formation. This model successfully explains a wide range of cosmological observations, including the near-flat geometry of the universe ($\Omega_{\text{total}} \approx 1$), and remains consistent with data from the cosmic microwave background (CMB), baryon acoustic oscillations (BAO), and supernovae.

However, despite its successes, the $\Lambda$CDM model faces significant theoretical and observational challenges. Chief among them is the so-called \textit{fine-tuning problem}: a severe discrepancy between the observed value of $\Lambda$ and predictions from quantum field theory. Additionally, the model struggles to reconcile different measurements of the Hubble constant ($H_0$), leading to the well-known \textit{Hubble tension}, and also discrepancies in the amplitude of matter clustering (the $S_8$ tension). Recent analyses by Planck \cite{Planck:2018vyg}, SH0ES \cite{Riess:2021jrx}, and DESI \cite{DESI:2024mwx} have highlighted the growing inconsistency between local and high-redshift determinations of $H_0$, while weak lensing surveys such as KiDS and DES \cite{Heymans:2021vmr, DES:2022ccp} reveal a lower amplitude of matter clustering, giving rise to the $S_8$ tension \cite{Nunes:2021ipq}. A comprehensive overview of these cosmological tensions and their possible origins is provided in the recent CosmoVerse White Paper \cite{CosmoVerseNetwork:2025alb}. These tensions indicate possible cracks in the standard cosmological model, motivating exploration beyond $\Lambda$CDM.
Furthermore, new BAO measurements from DESI and eBOSS \cite{DESI:2025zgx, eBOSS:2020abk} show a mild preference for models with dynamical dark energy or modified gravity, hinting that cosmic acceleration might not be perfectly consistent with a cosmological constant \cite{Arora:2025msq}. This strengthens the need to study extensions that allow late-time dynamics in the expansion history. In response to these issues, a variety of alternative models have been proposed, generally falling into three categories:

\begin{itemize}[label={$\star$}]
    \item \textbf{Dynamical Dark Energy Models:} These models replace the constant $\Lambda$ with a time-evolving component, typically represented by scalar fields. Examples include quintessence \cite{Tsujikawa:2013fta}, phantom fields \cite{Singh:2003vx}, $k$-essence \cite{Garcia:2012kr}, tachyon models \cite{Calcagni:2006ge}, and Chaplygin gas \cite{Gorini:2004by}. In such scenarios, the dark energy density evolves over cosmic time.

    \item \textbf{Modified Gravity Theories:} Instead of introducing a dark energy component, these models propose that gravity itself behaves differently on large scales. Modifications to General Relativity (GR), such as $f(R)$ gravity \cite{Sotiriou:2008rp}, $f(Q)$ gravity \cite{Heisenberg:2023lru}, $f(T)$ gravity \cite{Cai:2015emx, Kavya:2024ssu, Mishra:2024oln}, and scalar-tensor theories, attempt to account for cosmic acceleration through changes in the gravitational sector.

    \item \textbf{Vacuum-Based Models:} A distinctive alternative emerged between 1999 and 2004 with the development of Vacuum Cold Dark Matter (VCDM) models, particularly the ``Vacuum Metamorphosis'' scenario proposed by Parker and Raval. \cite{Parker:1999td, Parker:1999fc} In this framework, cosmic acceleration arises from quantum vacuum effects in curved spacetime. Specifically, a very light scalar field with mass $m \sim H_0$ remains inert at high curvature ($R \gg m^2$), but as the universe expands and the Ricci scalar $R$ decreases to the threshold $R \sim m^2$, non-perturbative vacuum effects activate. This triggers a phase transition, termed vacuum metamorphosis, leading to a sudden onset of late-time acceleration, without invoking a cosmological constant.
\end{itemize}

In contrast to early models that treated vacuum energy as a static quantity, VCDM-type theories propose that it is generated dynamically via quantum field interactions with curved spacetime. Modern realizations embed these ideas within scalar-tensor or modified gravity frameworks to gain additional theoretical flexibility and improve observational viability.  Particularly, the Type-II Minimally Modified Gravity (MMG) framework~\cite{DeFelice:2020eju,Scherer:2025esj} has recently attracted significant attention, as it links VCDM-like behaviour to a stable, covariant formulation that is compatible with current cosmological constraints. Such models provide a well-motivated route to interpret recent BAO and CMB indications of evolving late-time dynamics; indeed, a recent comprehensive analysis~\cite{Scherer:2025esj} reports strong ($\sim5\sigma$) evidence for a late-time dynamical dark-energy transition and shows that VCDM-like models give excellent fits to combined datasets. This body of work reinforces the motivation to study VCDM-inspired extensions, such as the Type-II MMG framework, as theoretically consistent and observationally viable alternatives to the standard $\Lambda$CDM picture. 

Various other generalizations have also been explored. These include single-field constructions with curvature-dependent effective masses~\citep{Onemli:2004mb,Doran:2006kp}, multi-field scenarios~\citep{Bento:2002ps}, and further developments within minimally modified gravity classes that aim to preserve stability and avoid ghost degrees of freedom~\citep{DeFelice:2020eju}. In this article, we concentrate on the observational consequences of these ideas: our primary goal is to test VCDM-inspired dynamics against current datasets rather than to develop new model-building techniques.

Unlike earlier studies that primarily examined the background expansion of VCDM-like or MMG models, the present work extends the analysis to include the growth of matter perturbations and their observational consequences. Our analysis is inspired by the VCDM framework originally introduced in Ref.~\cite{DeFelice:2020prd}, but it advances that work in both methodology and observational scope. We perform a comprehensive parameter estimation using substantially updated cosmological datasets, combining the covariance-corrected cosmic chronometer (CC) sample, the latest DESI DR2 baryon acoustic oscillation (BAO) measurements, the Union3 supernova compilation, and, importantly, the redshift-space distortion (RSD) data that directly probe the growth rate of cosmic structures. These additions enable a joint test of the VCDM scenario at both background and perturbation levels, providing a more complete and data-driven assessment of the model. In this framework, the parameter $\beta_H$ quantifies the deviation from the standard $\Lambda$CDM expansion rate, and the inclusion of high-precision DESI DR2 and RSD data significantly improves its constraints, reducing parameter degeneracies and allowing a stringent evaluation of the model's allowed departure from $\Lambda$CDM behavior. This integrated approach thus offers a more comprehensive observational test of the Type-II MMG framework and its VCDM-like dynamics in comparison with the standard cosmological model.

The article is organised in the following manner: The foundation of the background theory is presented in \autoref{vcdm}. The \autoref{datasets} is dedicated to a brief discussion on the concerned datasets. In \autoref{Results}, we present the results of this analysis, followed by concluding remarks in \autoref{conclusion}. 


\section{Background of VCDM Theory}
\label{vcdm}
To study the background cosmological dynamics, VCDM in the framework of type-II minimally modified gravity, we begin with the ADM decomposition of spacetime, which separates the spacetime metric into lapse function $N$, shift vector $N^i$, and spatial metric $\gamma_{ij}$ Within this decomposition, the gravitational sector of the theory is modified by introducing an auxiliary scalar field $\phi$, and a Lagrange multiplier $\lambda$ that enforces a modified constraint between the extrinsic curvature and the scalar field evolution.

The total Lagrangian density of the system, in terms of ADM variables and including the minimally modified sector, takes the form:
\begin{equation} 
 \mathcal{L}=N \sqrt{\gamma} \left[\frac{M_p^2}{2}\left(R+K_{ij}K^{ij}-K^2-2 V(\phi)-\frac{3}{4}\lambda^2-\lambda(K+\phi)-\lambda_{gf}^i\frac{1}{N}\partial_i \phi\right)+\mathcal{L_M} \right ]
\end{equation}
 Where \(M_p^2\) is the Planck's reduced mass, \(R\) is the Ricci scalar,
        \( K_{ij}\) is the extrinsic curvature of spatial slices,
       and \(K=\gamma^{ij}K_{ij}\) is the trace of extrinsic curvature and $\mathcal{L_M}$ represents the matter Lagrangian density describing the matter fields minimally coupled to the metric; The potential of the scalar field is denoted as \(V(\phi)\).
We express the spacetime metric $g_{\mu\nu}$ in terms of lapse, shift vector, and induced 3-metric $\gamma_{ij}$ on each hypersurface. The line element becomes:
\begin{equation}
ds^2 = -N^2 dt^2 + \gamma_{ij}(dx^i+N^idt)(dx^j+N^jdt)
\end{equation}
Under the assumption of a flat FLRW background, which describes a homogeneous and isotropic Universe, we consider the metric components as \(N = N(t)\), \(N^i = 0\), and \(\gamma_{ij} = a(t)^2 \delta_{ij}\). Following that, the action integral for the theory takes the form
\begin{equation}\label{eq:motaction}
    S = \int dt\, d^3x\, N a^3 M_P^2 \left[ -3 H^2 - V(\phi) - \frac{3 \lambda^2}{4} - \lambda (3 H + \phi) +\frac{\mathcal{L_M}}{M_P^2} \right]
\end{equation}
In the case of a perfect-fluid description, the matter Lagrangian coincides with the negative of the energy density when evaluated on the shell, hence $\mathcal{L_M}=-\rho$. Now equation \eqref{eq:motaction} becomes
\begin{equation} \label{eq:motactions}
    S = \int dt\, d^3x\, N a^3 M_P^2 \left[ -3 H^2 - V(\phi) - \frac{3 \lambda^2}{4} - \lambda (3 H + \phi) -\frac{\rho}{M_P^2} \right]
\end{equation}
where $\rho$ is the energy density of the matter component, obtained from the matter Lagrangian $\mathcal{L_M}$, \(H\) is the Hubble parameter, defined by \(H \equiv \frac{\dot{a}}{a N}\) and \(\rho\) denotes the energy density.
Taking the variation with respect to the auxiliary field $\lambda$ yields its equation of motion as
\begin{equation}\label{eq:motlambda}
    \lambda = -\frac{2}{3} \phi - 2 H.
\end{equation}
Similarly, taking the variation of the action with respect to the lapse function \(N\) yields its equation of motion, which corresponds to the first Einstein equation,
\begin{equation}\label{eq:motlapse}
    \phi^2 = 3 V(\phi) + \frac{3 \rho}{M_P^2}.
\end{equation}
Taking the variation of the action \eqref{eq:motactions} with respect to the scalar field $\phi$ and combining it with \eqref{eq:motlambda}, the corresponding equation of motion for field $\phi$ is
\begin{equation}\label{eq:motphi}
    \phi=\frac{3V_\phi}{2}-3H.
\end{equation}

In this setup, although the gravitational sector involves modifications through additional fields and couplings, the matter fields are assumed to be minimally coupled. As a result, they continue to obey the standard continuity (or conservation) equation, which for a homogeneous and isotropic universe takes the form:
\begin{equation}
\frac{\dot{\rho}}{N} + 3 H (\rho + P) = 0,
\end{equation}
where \( \rho \) and \( P \) denote the energy density and pressure of the matter sector, respectively. This equation reflects the local conservation of energy for the matter fields and remains unaltered by the modifications introduced in the gravitational sector.

Combining the motion equations \eqref{eq:motlapse} and \eqref{eq:motphi}, the Friedmann equation takes a more convenient form, 
\begin{equation}
3 M_P^2 H^2 = \rho + \rho_\phi,
\end{equation}
where the effective energy density associated with the scalar sector is given by
\begin{equation}
\rho_\phi = M_P^2 \left( V - \phi V_\phi + \frac{3}{4} V_\phi^2 \right).
\end{equation}
The potential \( V(\phi) \) is not uniquely determined, it can be reconstructed in terms of the e-folding variable \( \mathcal{N} \), which is defined through the scale factor as \( \mathcal{N} = \ln(a/a_0) \). The reconstructed form of the potential is given by \cite{DeFelice:2020eju}:
\begin{equation}
    V(\phi) = \frac{1}{3} \phi^2 - \frac{\rho\big(\mathcal{N}(\phi)\big)}{M_P^2},
\end{equation}
where \( \rho(\mathcal{N}) \) denotes the energy density expressed as a function of \( \mathcal{N} \). For a fixed form of the Hubble parameter \( H(z) \), the shape of the potential depends on the arbitrary choice of the initial value of the scalar field \( \phi \). Changing this initial value results in a shift of the potential by a constant and adds a term linear in \( \phi \). However, the potential itself does not influence the dynamics at the level of the background or linear perturbation theory, which are determined entirely by \( H(z) \) and its derivatives. Moreover, the VCDM theory reduces to General Relativity whenever \( V_{\phi\phi} = 0 \).

Here we adopt the Hubble function from \cite{DeFelice:2020cpt}, which reads
\begin{equation}\label{eq:H1}
H^2={H_\Lambda}^2+A_1 {H_0}^2\left\{1-\text{tanh}\left(\frac{z-A_2}{A_3}\right)\right\},
\end{equation}
where ${H_\Lambda}^2 \equiv {H_{\Lambda_0}}^2 \left( \tilde\Omega_\Lambda + \tilde\Omega_{m0} (1+z)^3 + \tilde\Omega_{r0} (1+z)^4 \right)$ is the Friedmann equation for the flat $\Lambda$CDM model. Here $A_1$, $A_2$, and $A_3$ are the independent parameters and $\tilde\Omega_\Lambda+\tilde\Omega_{m0}+\tilde\Omega_{r0}=1$. Considering the present time scenario in \eqref{eq:H1}, the parameter $A_1$ can be obtained as 
\begin{equation}
    A_1=\left(1-\frac{{H_{\Lambda0}}^2}{{H_0}^2}\right)\left(1+\text{tanh}\left(\frac{A_2}{A_3}\right)\right)^{-1}.
\end{equation}
Additionally, the following parameter redefinitions are adopted:
\[
\tilde{\Omega}_{m0} = \Omega_{m0} \frac{{H_0}^2}{{H_{\Lambda 0}}^2}, \quad 
\tilde{\Omega}_{r0} = \Omega_{r0} \frac{{H_0}^2}{{H_{\Lambda 0}}^2}, \quad
\beta_H = \frac{H_{\Lambda 0}}{H_0},
\]
to obtain the final expression for the Hubble parameter:
\begin{equation}
\frac{H^2}{{H_0}^2} =
\beta_H^2+\Omega_{m0} \left((1+z)^3 - 1\right) +
\Omega_{r0} \left((1+z)^4 - 1\right) +
(1 - \beta_H^2) \left( \frac{1 + \tanh\left( \frac{A_2 - z}{A_3} \right)}{1 + \tanh\left( \frac{A_2}{A_3} \right)} \right).
\end{equation}

To constrain the parameters appearing in the above expression, we employ a statistical parameter estimation technique, the details of which will be discussed in the subsequent sections.

\section{Datasets}\label{datasets}

In this work, we adopt a combination of complementary cosmological probes that together provide a comprehensive and internally consistent description of the late-time expansion history and structure growth. The selection of datasets is guided by two primary considerations: (i) the need for model-independent and minimally correlated observables, and (ii) the requirement of compatibility among the most recent and homogeneous measurements. Specifically, we employ the covariance-corrected CC compilation~\cite{Moresco:2020fbm} to directly determine the Hubble parameter \(H(z)\) without assuming a fiducial cosmology, the BAO measurements from DESI Data Release~2 (DR2)~\cite{DESI:2024mwx} to anchor the distance scale with unprecedented precision, and the RSD data~\cite{Alestas:2022gcg} to constrain the linear growth rate of cosmic structures and the parameter \(S_8\). In addition, the Union3 Type~Ia supernova sample~\cite{Rubin:2023ovl} provides luminosity distance information across a broad redshift range, ensuring an accurate mapping of the cosmic expansion history. 

This combination of datasets offers an optimal balance between precision and consistency: the CC data constrain the differential expansion rate, DESI DR2 and Union3 probe the integrated distance measures, and RSD captures the growth of matter perturbations, thereby allowing a joint and robust evaluation of the cosmological model. Furthermore, we do not add the PantheonPlus supernova compilation due to its well-documented $>3\sigma$ tension with the considered datasets, which would introduce bias and degrade the consistency of the joint constraints. The chosen dataset ensemble thus represents the most statistically coherent and up-to-date combination currently available for testing minimally modified gravity models.

\subsection{Cosmic Chronometers}

Cosmic chronometers provide a direct and model-independent technique for measuring the Hubble parameter, \(H(z)\), by utilizing the differential age evolution of passively evolving galaxies~\cite{Jimenez:2001gg}. The basic principle relies on the fact that, for galaxies evolving without significant star formation, the derivative of redshift with respect to cosmic time can be related to the expansion rate as
\begin{equation}
H(z) = -\frac{1}{1+z}\frac{\mathrm{d}z}{\mathrm{d}t}.
\end{equation}
This approach, based on the spectroscopic dating of stellar populations, avoids assumptions about integrated distance indicators, such as standard candles or rulers. Consequently, it serves as a purely differential probe of the expansion history, offering crucial constraints on late-time cosmological evolution and tests of modified gravity.

In the present analysis, we employ an updated compilation of 34 CC measurements spanning the redshift range \(0 < z < 2\), drawn from multiple surveys as summarized in Refs.~\cite{Moresco:2016mzx,Moresco:2020fbm}. Among these, 15 data points correspond to the homogeneous subset provided by Moresco et~al.~(2020), for which a full covariance matrix is publicly available\footnote{\url{https://gitlab.com/mmoresco/CCcovariance}}. These correlated points incorporate systematic uncertainties arising from stellar population synthesis modeling, metallicity effects, and calibration procedures. The remaining 19 measurements, collected from independent analyses, are treated as uncorrelated and included with their individual statistical errors. 

To ensure consistency, our total chi-square for the CC dataset is constructed by combining the correlated and uncorrelated contributions as \cite{kavya:2025sr}
\begin{equation}
\chi^2_{\mathrm{CC}} = \chi^2_{\mathrm{corr}} + \chi^2_{\mathrm{uncorr}},
\end{equation}
where the correlated term accounts for the covariance matrix of the homogeneous subset:
\begin{equation}
\chi^2_{\mathrm{corr}} =
\Delta \mathbf{H}_{\mathrm{corr}}^{T} \,
C_{\mathrm{corr}}^{-1} \,
\Delta \mathbf{H}_{\mathrm{corr}},
\end{equation}
and the uncorrelated part is expressed as
\begin{equation}
\chi^2_{\mathrm{uncorr}} =
\sum_{i=1}^{N_{\mathrm{uncorr}}}
\frac{\left[H_{\mathrm{th}}(z_i) - H_{\mathrm{obs}}(z_i)\right]^2}{\sigma_{H,i}^2},
\end{equation}
with \(N_{\mathrm{uncorr}} = 19\). Here, 
\(\Delta \mathbf{H}_{\mathrm{corr}} = \mathbf{H}_{\mathrm{th}} - \mathbf{H}_{\mathrm{obs}}\) 
represents the residual vector for the correlated subset, and 
\(C_{\mathrm{corr}}\) denotes the corresponding covariance matrix from Ref.~\cite{Moresco:2020fbm}. 

This covariance-corrected treatment significantly improves the reliability of the inferred cosmological parameters by properly accounting for correlated systematics, thereby providing a consistent and robust comparison with the DESI DR2, Union3, and RSD datasets used in this work.

\subsection{DESI DR2}
In this work, we make use of the BAO measurements from the Dark Energy Spectroscopic Instrument (DESI) DR2, which provides the most extensive sample to date---comprising over 14 million galaxies and quasars distributed across seven redshift bins within the range \(0.295 \leq z \leq 2.33\)~\cite{DESI:2024mwx}. The DR2 compilation includes four primary tracers: the bright galaxy sample (BGS)~\cite{Hahn:2022dnf}, luminous red galaxies (LRG)~\cite{DESI:2022gle}, emission line galaxies (ELG)~\cite{Raichoor:2022jab}, and quasars (QSO)~\cite{Chaussidon:2022pqg}, all of which feature significantly improved statistical precision compared to DR1.

The BGS measurements primarily provide isotropic BAO constraints at well-defined effective redshifts, whereas the ELG, LRG, and QSO samples deliver both isotropic and anisotropic determinations of cosmological distances. In particular, the anisotropic analyses yield constraints on the transverse comoving distance \(D_M / r_d\), the Hubble distance \(D_H / r_d\), and the volume-averaged distance \(D_V / r_d\), with the isotropic subsets offering stringent bounds on \(D_V / r_d\).

The relevant theoretical relations used in our analysis are expressed as
\begin{gather}
r_d = \frac{1}{H_0} \int_{z_d}^{\infty} \frac{c_s(z')}{E(z')} \, dz', \\
D_M(z) = \frac{c}{H_0} \int_0^z \frac{dz'}{E(z')}, \\
D_H(z) = \frac{c}{H(z)}, \\
D_V(z) = \left[z \, D_M^2(z) \, D_H(z)\right]^{1/3},
\end{gather}
where \(c_s\) denotes the sound speed of the baryon--photon fluid and \(c\) is the speed of light.

The total chi-square function used for the DESI likelihood is expressed as
\begin{equation}
\chi^2_{\text{DESI}} = \chi^2_{\text{iso}} + \chi^2_{\text{aniso}},
\end{equation}
where \(\chi^2_{\text{iso}}\) and \(\chi^2_{\text{aniso}}\) represent the contributions from isotropic and anisotropic BAO measurements, respectively.

The isotropic term is given by
\begin{equation}
\chi^2_{\text{iso}} = \sum_{i} \frac{\left[A_{\text{th}}(z_i) - A_{\text{obs}}(z_i)\right]^2}{\sigma_A^2(z_i)},
\end{equation}
where \(A = D_V / r_d\), and \(\sigma_A(z_i)\) denotes the observational uncertainty at each redshift.

For the anisotropic component, the chi-square is constructed as
\begin{equation}
\chi^2_{\text{aniso}} = \Delta B^{T} \, C_{\text{DESI}}^{-1} \, \Delta B,
\end{equation}
where \(\Delta B\) is the vector of residuals corresponding to the theoretical and observed values of \(D_H / r_d\) and \(D_M / r_d\), and \(C_{\text{DESI}}\) represents the covariance matrix associated with the correlated DESI DR2 data~\cite{DESI:2025zgx}.

\subsection{Redshift-Space Distortion}

The incorporation of RSD measurements provides valuable information on the growth rate of cosmic structures. This phenomenon arises because peculiar velocities of galaxies distort their apparent clustering when mapped in redshift space, leading to anisotropies in the observed large-scale structure. On large scales, overdense regions appear compressed along the line of sight due to coherent infall motions, whereas on small scales, random motions within virialized structures produce elongations, commonly referred to as the ``Finger-of-God'' effect. These combined effects modify the two-point correlation function and generate an anisotropic power spectrum. In our analysis, we include RSD data jointly with DESI and other datasets to compare the theoretical prediction of the present-day quantity \({S_8}_0\) with its observational estimate.

When converting redshift to distance in RSD analyses, the adoption of a fiducial cosmology introduces additional anisotropies known as the Alcock--Paczynski (AP) effect. To mitigate the bias introduced by this effect, we apply a correction factor following Ref.~\cite{Macaulay:2013swa}. The corrected growth rate can then be written as
\begin{equation}
f\sigma_8(a) \approx 
\frac{H_{\mathrm{model}}(a) D_{A,\mathrm{model}}(a)}
{H_{\mathrm{fid}}(a) D_{A,\mathrm{fid}}(a)} 
f\sigma_{8,\mathrm{fid}}(a),
\end{equation}
where the subscripts ``fid'' and ``model'' denote quantities corresponding to the fiducial \(\Lambda\)CDM cosmology and the tested model, respectively. Here, \(H(a)\) and \(D_A(a)\) represent the Hubble parameter and the angular diameter distance as functions of the scale factor.

The chi-square estimator used for the RSD likelihood is defined as
\begin{equation}
\chi^2_{\mathrm{RSD}} = 
\sum_{i,j=1}^{20} 
\left[f\sigma_8^{\mathrm{th}}(z_i) - f\sigma_8^{\mathrm{obs}}(z_i)\right]
\left(C^{-1}\right)_{ij}
\left[f\sigma_8^{\mathrm{th}}(z_j) - f\sigma_8^{\mathrm{obs}}(z_j)\right],
\end{equation}

\noindent
where \(f\sigma_8^{\mathrm{th}}(z_i)\) and \(f\sigma_8^{\mathrm{obs}}(z_i)\) denote the theoretical and observed growth rate values at redshift \(z_i\), respectively, and \(C^{-1}_{ij}\) is the inverse covariance matrix that accounts for correlations among the data points. The updated compilation of the RSD measurements used in this work is provided in Table~II of Ref.~\cite{Alestas:2022gcg}.

\subsection{Union3}
The Union3 supernova compilation incorporates a total of 2087 Type Ia supernovae (SNe Ia), sourced from 24 distinct observational surveys. Each event in this sample has been homogenized using the SALT3 light-curve fitting method, allowing for consistent calibration of luminosity indicators such as stretch and color. To facilitate accurate statistical inference, the dataset undergoes processing via the UNITY1.5 Bayesian analysis framework, which systematically addresses selection biases, intrinsic dispersion, and systematic uncertainties. This methodology yields 22 binned estimates of the supernova distance modulus, spanning a redshift range from $z=0.05$ to $z=2.26$. These binned observations effectively trace the expansion history of the Universe over a broad redshift interval and serve as a crucial empirical input for the cosmographic analysis and theoretical model evaluation presented in this work \citep{Rubin:2023ovl}.

The chi-square function for the Union3 supernova compilation is constructed using the observed and theoretical distance moduli as
\begin{equation}
\chi^2_{\mathrm{SN}} = 
\Delta \boldsymbol{\mu}^{T} 
\, C_{\mathrm{SN}}^{-1} \,
\Delta \boldsymbol{\mu},
\end{equation}
where 
\(\Delta \boldsymbol{\mu} = \boldsymbol{\mu}_{\mathrm{th}} - \boldsymbol{\mu}_{\mathrm{obs}}\)
denotes the residual vector between the theoretical and observed distance modulus values, and 
\(C_{\mathrm{SN}}\) is the covariance matrix associated with the Union3 binned dataset~\cite{Rubin:2023ovl, Mishra:2025hark}. 
The theoretical distance modulus is defined as
\begin{equation}
\mu_{\mathrm{th}}(z) = 5 \log_{10} \left[ \frac{D_L(z)}{\mathrm{Mpc}} \right] + 25,
\end{equation}
where the luminosity distance is given by 
\begin{equation}
D_L(z) = (1+z)\frac{c}{H_0}\int_0^z \frac{dz'}{E(z')}.
\end{equation}

\section{Methodology and Results}\label{Results}
The Markov Chain Monte Carlo (MCMC) technique is employed to perform parameter estimation within a Bayesian framework. MCMC generates samples from the posterior distribution of the model parameters by constructing a Markov chain whose equilibrium distribution corresponds to the desired posterior. We utilize the affine-invariant ensemble sampler implemented in the \texttt{emcee} Python package \cite{Foreman-Mackey:2012any}, which is well-suited for exploring complex, multidimensional parameter spaces.

In this analysis, the likelihood function is constructed based on the chi-squared ($\chi^2$) statistic, which quantifies the difference between the model predictions and observational data. For a dataset with observed values $y_i^{\text{obs}}$, theoretical predictions $y_i^{\text{th}}(\boldsymbol{\theta})$ depending on parameters $\boldsymbol{\theta}$, and associated uncertainties $\sigma_i$, the chi-squared is defined as:

\begin{equation}
    \chi^2(\boldsymbol{\theta}) = \sum_{i} \left( \frac{y_i^{\text{obs}} - y_i^{\text{th}}(\boldsymbol{\theta})}{\sigma_i} \right)^2.
\end{equation}

For correlated datasets, such as those with a non-diagonal covariance matrix $C$, the generalized form of the chi-squared becomes:

\begin{equation}
    \chi^2(\boldsymbol{\theta}) = \Delta \mathbf{y}^T C^{-1} \Delta \mathbf{y},
\end{equation}
where $\Delta \mathbf{y} = \mathbf{y}^{\text{obs}} - \mathbf{y}^{\text{th}}(\boldsymbol{\theta})$ is the residual vector and $C^{-1}$ is the inverse covariance matrix.

The likelihood function is related to the chi-squared by:

\begin{equation}
    \mathcal{L}(\boldsymbol{\theta}) \propto \exp\left( -\frac{1}{2} \chi^2(\boldsymbol{\theta}) \right).
\end{equation}

The posterior distribution is then given by Bayes' theorem:

\begin{equation}
    \mathcal{P}(\boldsymbol{\theta} | \text{data}) \propto \mathcal{L}(\boldsymbol{\theta}) \, \pi(\boldsymbol{\theta}),
\end{equation}
where $\pi(\boldsymbol{\theta})$ is the prior distribution of the parameters. After an initial burn-in period, the remaining MCMC samples are used to derive marginalized constraints and credible intervals for the parameters.

\begin{figure}
    \centering
    \includegraphics[width=0.75\linewidth]{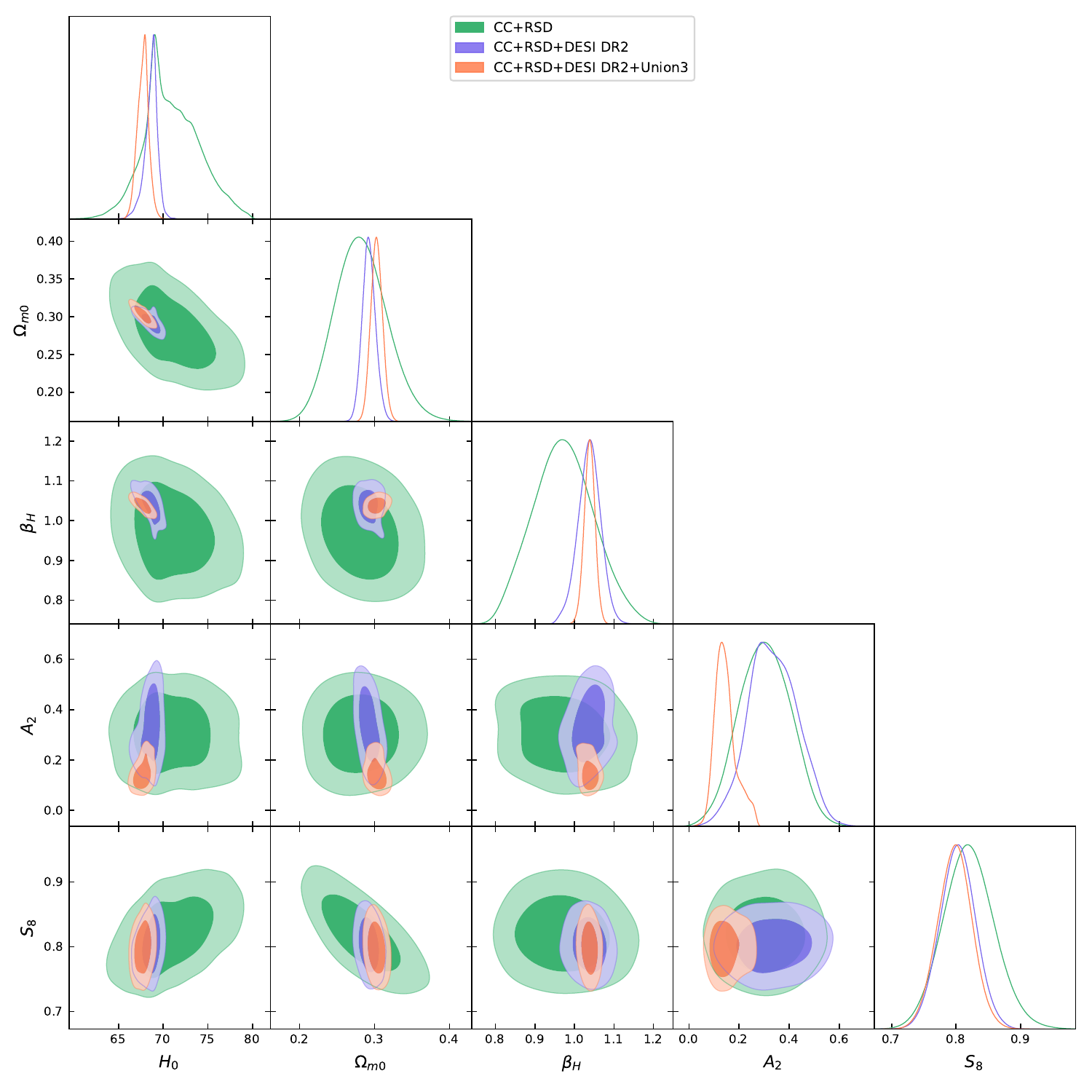}
    \caption{Upto $2\sigma$ ($95\%$ CL) 2D likelihood contours for the parameter space $\{H_0,\Omega_{m0},\beta_H,A_2,S_8\}$.}
    \label{fig:con}
\end{figure}

\begin{figure}
    \centering
    \includegraphics[width=0.6\linewidth]{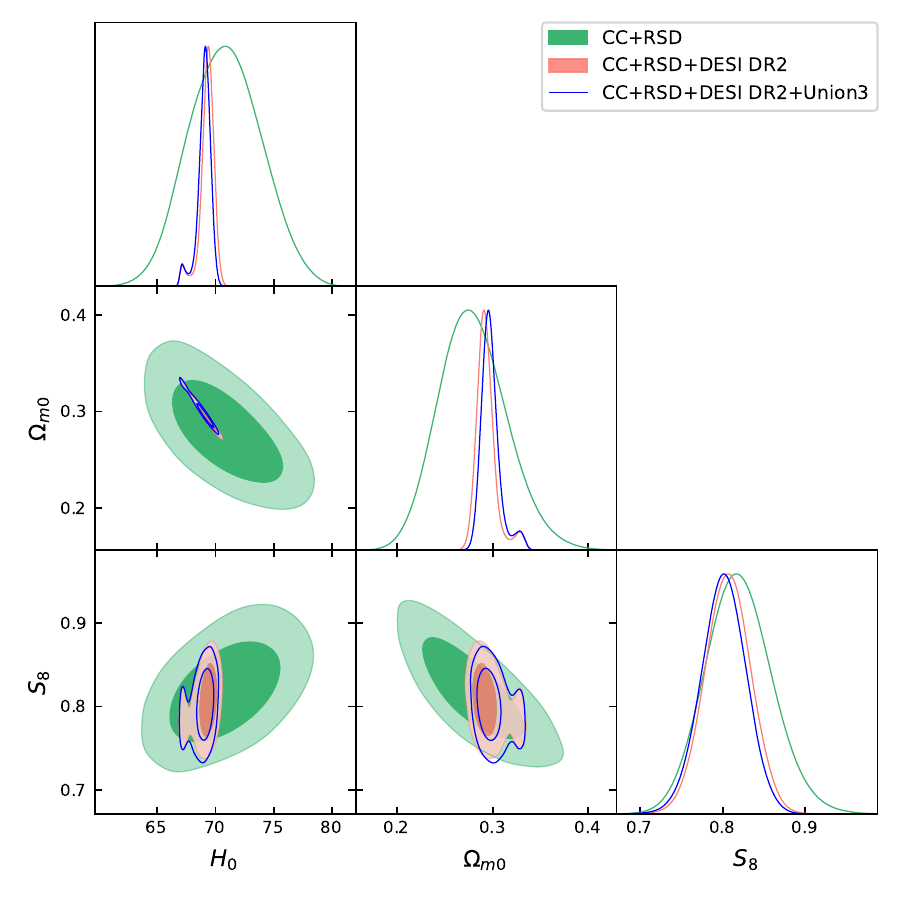}
    \caption{Upto $2\sigma$ ($95\%$ CL) 2D likelihood contours for $\Lambda$CDM.}
    \label{fig:lcdmcon}
\end{figure}

The minimally modified VCDM framework serves as the underlying cosmological model in this analysis. We perform the MCMC simulations with combinations of CC, DESI DR2, RSD, and Union3 datasets. Upto $2\sigma$ contours are presented in \autoref{fig:con}, where the inner dark shade corresponds to $1\sigma$ ($68\%$ confidence level) and the lighter is $2\sigma$ ($95\%$ confidence level). The resulting $1\sigma$ values are summarized in the \autoref{tab:tab1}. For a consistent comparison, we have performed the MCMC analysis for the standard $\Lambda$CDM model using the same dataset combinations as adopted for VCDM. The corresponding results are illustrated in \autoref{fig:lcdmcon} and summarized in \autoref{tab:lcdm}.

\begin{figure}
    \centering
    \includegraphics[width=0.6\linewidth]{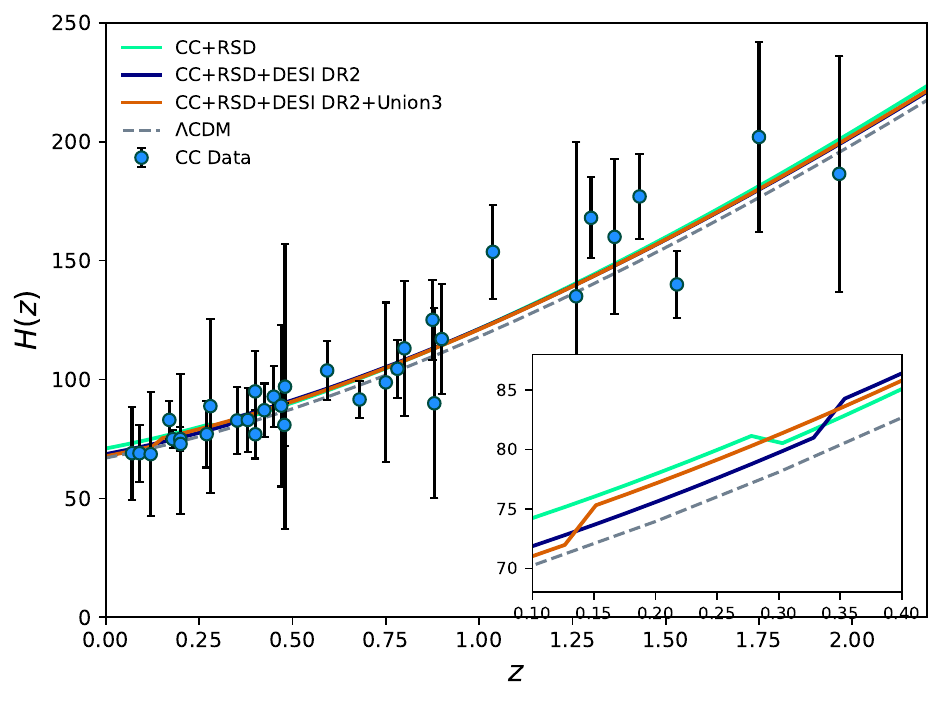}
    \caption{Hubble parameter against redshift along with the CC data. }
    \label{fig:Hz}
\end{figure}

In \autoref{fig:Hz}, we present the evolution of the Hubble parameter as a function of redshift for various dataset combinations within the VCDM framework, alongside the corresponding $\Lambda$CDM prediction for comparison. The 34 cosmic chronometer (CC) measurements with their respective error bars are also plotted to illustrate the agreement with observational data. Consistent with the findings of De~Felice~et~al.~\cite{DeFelice:2020cpt}, the VCDM model exhibits a distinct transition in the Hubble expansion rate around the redshift \(z \simeq 0.3\), which corresponds to the value of the model parameter \(A_2\). The sharpness or width of this transition is governed by the parameter \(A_3\), characterizing how rapidly the expansion rate changes near the transition epoch. To highlight this behavior, the region \(z \in (0.1, 0.4)\) has been magnified in the inset of \autoref{fig:Hz}, where the deviation of the VCDM curves from the smooth $\Lambda$CDM evolution becomes clearly visible. The transition feature represents the dynamical effect of vacuum metamorphosis in the VCDM scenario, marking the onset of the late-time acceleration phase.

The deceleration parameter, \( q(z) \), serves as a key kinematic quantity in cosmology, offering insights into the accelerating or decelerating nature of the Universe's expansion. Defined as \( q(z) = -1 - \frac{\dot{H}}{H^2} \), it directly measures the rate of change of the Hubble parameter. A negative value of \( q \) indicates accelerated expansion, while a positive value corresponds to deceleration. Precise reconstruction of \( q(z) \) across different redshifts allows for stringent tests of cosmological models and provides crucial information about the underlying dynamics of dark energy. Moreover, the evolution of \( q(z) \) plays a fundamental role in identifying the transition epoch from deceleration to acceleration, which is essential for understanding the large-scale evolution of the Universe. 

\begin{figure}
    \centering
    \includegraphics[width=0.6\linewidth]{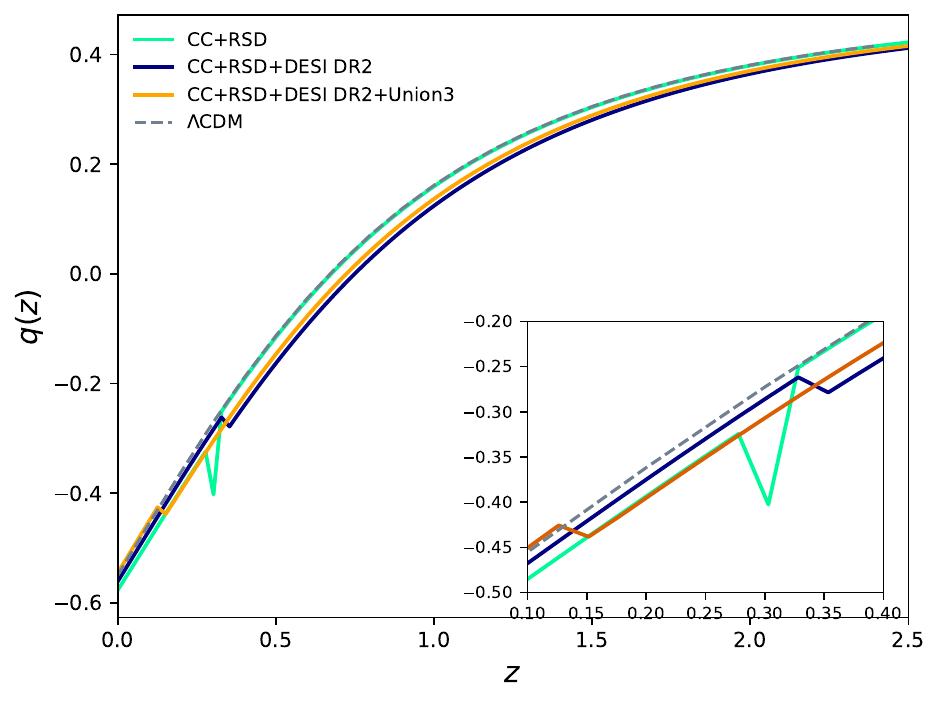}
    \caption{Evolution of the Universe through the Deceleration parameter.}
    \label{fig:qz}
\end{figure}

In \autoref{fig:qz}, we display the redshift evolution of the deceleration parameter for the VCDM model across the three dataset combinations, along with the corresponding $\Lambda$CDM prediction. The results reveal that all realizations of the VCDM framework exhibit a smooth transition from a decelerating to an accelerating phase, consistent with observational expectations. The transition redshift, \(z_t\), marking the epoch where \(q(z_t) = 0\), lies within the range \(0.67 \lesssim z_t \lesssim 0.75\) for all combinations, in excellent agreement with the standard $\Lambda$CDM value (\(z_t \approx 0.67\)). The present-day value of the deceleration parameter is found to be \(q_0 \approx -0.56\), indicating an excellent match to $\Lambda$CDM. These results are summarized in \autoref{tab:tabq0}, which also lists the corresponding transition redshifts for each combination. 

\begin{figure}
    \centering
    \includegraphics[width=0.6\linewidth]{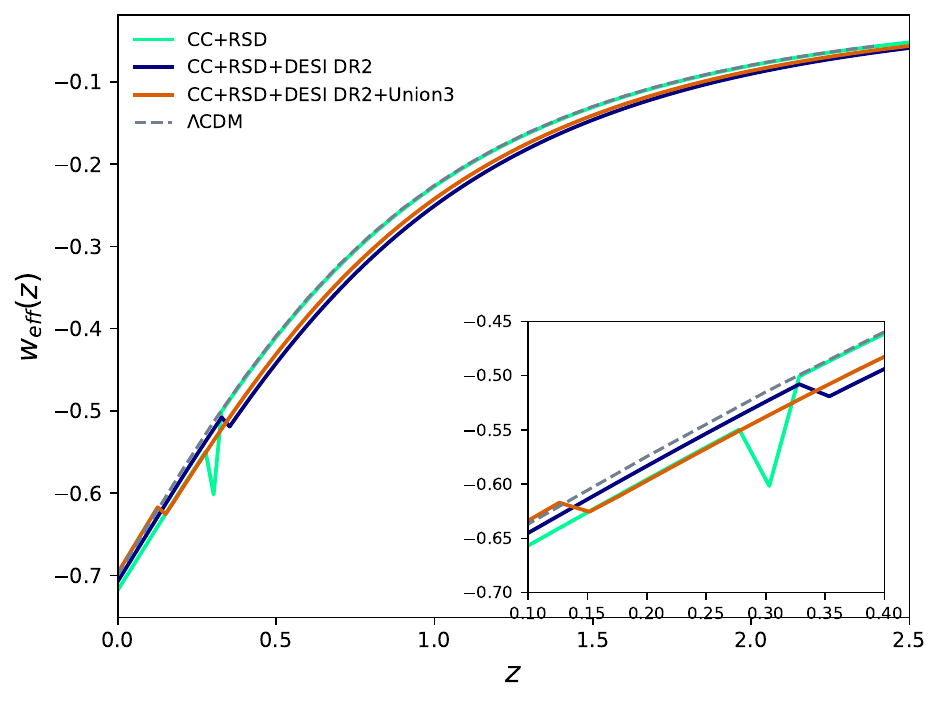}
    \caption{Evolution of the Universe through effective EoS parameter.}
    \label{fig:wz}
\end{figure}

The effective equation of state (EoS) parameter, \( w_{\text{eff}}(z) \), provides a comprehensive description of the total cosmic fluid driving the Universe's expansion. It is defined as \( w_{\text{eff}}(z) = -1 + \frac{2}{3}(1+z)\frac{H'(z)}{H(z)} \), relating the Hubble parameter's evolution to the dynamical properties of the cosmic medium. Unlike specific dark energy models, \( w_{\text{eff}}(z) \) encapsulates the collective behavior of all energy components, making it a valuable tool for model-independent diagnostics. Its evolution traces the transition from matter domination (\( w_{\text{eff}} \approx 0 \)) to the late-time acceleration (\( w_{\text{eff}} < -1/3 \)), offering crucial insights into the cosmic acceleration mechanism and the viability of competing cosmological models. 
 As shown in \autoref{fig:wz}, all three VCDM realizations follow the expected high-redshift behavior \(w_{\mathrm{eff}} \rightarrow 0\), corresponding to the matter-dominated epoch, and subsequently transition toward negative values at late times, indicating accelerated expansion. The present-day values \(w_{\mathrm{eff},0}\) listed in \autoref{tab:tabq0} are close to \(-0.7\), consistent with the standard $\Lambda$CDM expectation. 

The transition in \(w_{\mathrm{eff}}(z)\) around \(z \sim 0.3\)-0.4 corresponds to the same epoch identified in the Hubble evolution (see \autoref{fig:Hz}) and in the deceleration parameter \(q(z)\), reinforcing the internal consistency of the model. The inset panel highlights this transition region, clearly showing that the VCDM model yields a mildly dynamical equation of state that approaches $\Lambda$CDM.

 The behavior and numerical values of these parameters are also found to be consistent with the results reported in \cite{Mishra:2025rhi}, further supporting the robustness of our analysis.

\begin{figure}
    \centering
    \includegraphics[width=0.6\linewidth]{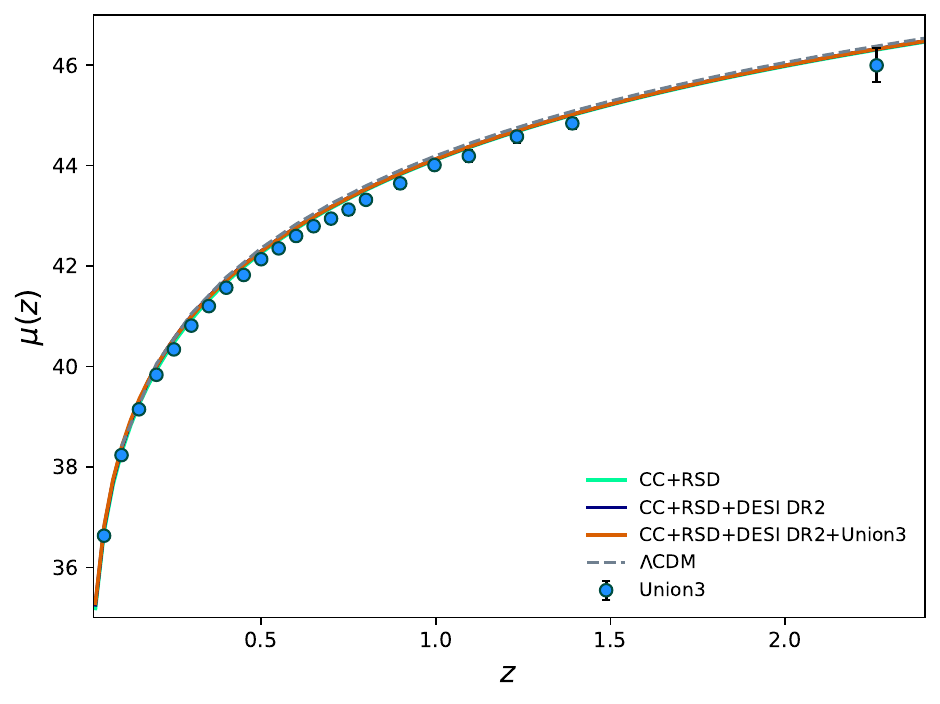}
    \caption{Distance modulus against redshift along with the Union3 data.}
    \label{fig:muz}
\end{figure}

Furthermore, we present the distance modulus profile in \autoref{fig:muz}, plotted against the Union3 data with corresponding error bars. It is evident that the combinations provide a good fit, showing excellent agreement with both the \(\Lambda\)CDM model and the observational data.

\begin{figure}
    \centering
    \includegraphics[width=0.6\linewidth]{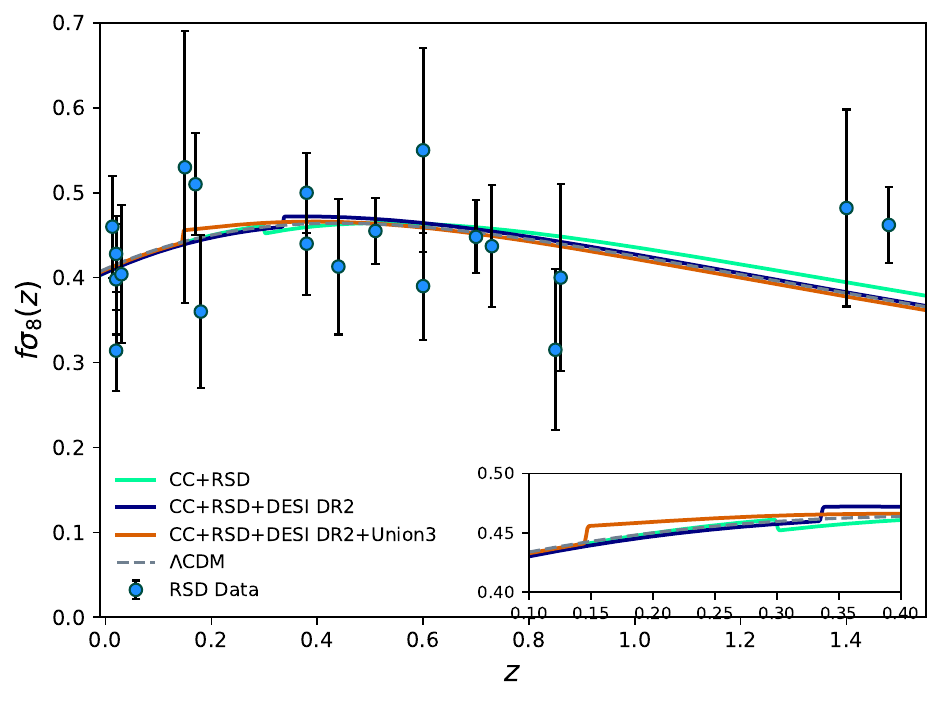}
    \caption{Evolution of the growth rate of cosmic structures, expressed as \(f\sigma_8(z)\), for the VCDM model across different dataset combinations, compared with the standard $\Lambda$CDM prediction and the latest RSD measurements.}
    \label{fig:fs8}
\end{figure}

To further test the model beyond background dynamics, we examine the growth rate of cosmic structures through the observable \(f\sigma_8(z)\), as shown in \autoref{fig:fs8}. This quantity, directly measured from redshift-space distortion (RSD) observations, combines the linear growth rate \(f = \frac{d\ln D}{d\ln a}\) with the rms amplitude of matter fluctuations, \(\sigma_8\). The evolution of \(f\sigma_8(z)\) provides a sensitive probe of the underlying gravity theory and allows us to constrain the growth parameter \(S_8 = \sigma_8\sqrt{\Omega_m/0.3}\).

In \autoref{fig:fs8}, the VCDM model predictions for all three dataset combinations are compared with the RSD data and the corresponding $\Lambda$CDM curve. The results demonstrate excellent consistency between the VCDM model and the observed growth measurements across the entire redshift range \(0 < z < 1.5\). The inclusion of DESI DR2 data leads to a slightly reduced growth amplitude, bringing the model predictions into even closer agreement with the RSD observations. Notably, the full combination (CC+RSD+DESI DR2+Union3) yields a present-day value of \(S_8 \simeq 0.80\) (see \autoref{fig:con} and \autoref{tab:tab1}), which is compatible with weak-lensing constraints and marginally lower than the Planck-inferred value. This indicates that the VCDM model can reproduce both the expansion history and the growth of cosmic structures without introducing significant tension with current observations.

\subsection{Model Comparison and Statistical Criteria}

To quantitatively assess the relative performance of the VCDM model with respect to the standard $\Lambda$CDM scenario, we evaluate the minimum $\chi^2$ values for each dataset combination, along with the corresponding Akaike Information Criterion (AIC) and Bayesian Information Criterion (BIC). The $\chi^2$ statistic provides a direct measure of the goodness-of-fit, while AIC and BIC offer information-theoretic metrics that penalize models with additional parameters, thereby balancing fit quality against model complexity~\cite{Akaike:1974,Burnham:2002,BIC:1978}.

The AIC and BIC are defined respectively as
\begin{equation}
\mathrm{AIC} = \chi^2_{\mathrm{min}} + 2k, \qquad 
\mathrm{BIC} = \chi^2_{\mathrm{min}} + k \ln N,
\end{equation}
where $k$ denotes the number of free parameters in the model and $N$ represents the number of data points used in the fit. The relative performance between two competing models is then quantified by the differences $\Delta\mathrm{AIC}$ and $\Delta\mathrm{BIC}$, with negative values indicating statistical preference for the alternative model.

Table~\ref{tab:chi2} lists the minimum $\chi^2$ values obtained for both the VCDM and $\Lambda$CDM models across different dataset combinations, while Table~\ref{tab:aicbic} presents the corresponding AIC and BIC values. The results demonstrate that the VCDM model consistently achieves a better fit than $\Lambda$CDM, with $\Delta\chi^2_{\mathrm{min}} < 0$ in all cases. Notably, when the full dataset combination (CC+RSD+DESI DR2+Union3) is employed, the VCDM model yields a substantial improvement of $\Delta\chi^2_{\mathrm{min}} = -10.372$ relative to $\Lambda$CDM. The AIC and BIC comparisons indicate that VCDM remains competitive even after accounting for model complexity, exhibiting lower or comparable information criterion values in the most comprehensive dataset combination.

Overall, these results suggest that the VCDM framework provides a slightly better description of current observational data compared to $\Lambda$CDM, particularly when late-time geometric and growth measurements are jointly considered. These findings collectively indicate that VCDM provides a viable and statistically consistent extension of the standard cosmological model, capable of reproducing both background and growth-level observables with minimal additional complexity.

\begin{table}
\centering
\small 

\begin{tabular}{|c|c|c|c|c|c|}
\hline
Dataset & $H_0$ & $\Omega_{m0}$ & $\beta_H$ & $A_2$ & $S_8$ \\
\hline \hline
CC+RSD & $71.0\pm3.1$ & $0.282_{-0.037}^{+0.032}$ & $0.972\pm0.077$ & $0.30 \pm 0.10$ & $0.820\pm 0.040$ \\ 
CC+RSD+DESI DR2 & $68.63_{-0.54}^{+0.77}$ & $0.2925 \pm 0.0089$ & $1.037\pm0.029$ & $0.336 \pm 0.098$ & $0.803\pm0.028$ \\ 
CC+RSD+DESI DR2+Union3 & $67.72\pm0.63$ & $0.3028 \pm 0.0078$ & $1.038\pm0.014$ & $0.146_{-0.051}^{+0.026}$ & $0.798\pm0.027$ \\
\hline
\end{tabular}

\caption{The numerical $1\sigma$ values of the parameters for the VCDM model.}
\label{tab:tab1}
\end{table}

\begin{table}
\centering
\small 

\begin{tabular}{|c|c|c|c|}
\hline
Dataset & $H_0$ & $\Omega_{m0}$ & $S_8$ \\
\hline \hline
CC+RSD & $70.8\pm3.1$ & $0.279_{-0.038}^{+0.032}$ & $0.820\pm 0.040$ \\ 
CC+RSD+DESI DR2 & $69.25_{-0.38}^{+0.64}$ & $0.2936_{-0.011}^{+0.0062}$ & $0.806\pm0.028$ \\ 
CC+RSD+DESI DR2+Union3 & $69.01_{-0.36}^{+0.62}$ & $0.2980_{-0.011}^{+0.0061}$ & $0.801\pm0.027$ \\
\hline
\end{tabular}

\caption{The numerical $1\sigma$ values of the parameters for the $\Lambda$CDM model.}
\label{tab:lcdm}
\end{table}

\begin{table}
    \centering
\small 

    \begin{tabular}{|c|c|c|c|}
    \hline
       Dataset/Model  &  $q_0$ & $z_t$ & $w_0$\\
       \hline \hline
     CC+RSD  &  $-0.5770$ & $0.6752$ & $-0.7180$  \\ 
     CC+RSD+DESI DR2  &  $-0.5613$ & $0.7493$ & $-0.7075$  \\ 
     CC+RSD+DESI DR2+Union3  &  $-0.5458 $ & $0.7232$ & $-0.6972$  \\
     $\Lambda$CDM  &  $-0.55 $ & $0.672$ & $-0.7$  \\

     \hline
    \end{tabular}
    
    \caption{The transition redshifts and present values of $q(z)$ and $w_{eff}(z)$.}
    \label{tab:tabq0}
\end{table}

\begin{table}
\centering
\small 

\begin{tabular}{|c|c|c|c|}
\hline
Combination & $\chi_{min}^2(VCDM)$ & $\chi_{min}^2(\Lambda CDM)$ & $\Delta\chi_{min}^2$ \\
\hline \hline
CC+RSD & $29.268$ & $30.905$ & $-1.637$ \\ 
CC+RSD+DESI DR2 & $41.284$ & $47.027$ & $-5.743$ \\ 
CC+RSD+DESI DR2+Union3 & $67.977$ & $78.349$ & $-10.372$ \\
\hline
\end{tabular}

\caption{The obtained minimum $\chi^2$ for VCDM and $\Lambda$CDM.}
\label{tab:chi2}
\end{table}
\begin{table}
\centering
\footnotesize
\setlength{\tabcolsep}{5pt}

\begin{tabular}{|c|cc|c|cc|c|}
\hline
Dataset & \multicolumn{2}{c|}{AIC} & $\Delta$AIC & \multicolumn{2}{c|}{BIC} & $\Delta$BIC \\ 
\cline{2-3}\cline{5-6}
 & VCDM & $\Lambda$CDM &  & VCDM & $\Lambda$CDM &  \\
\hline\hline
CC+RSD & 39.268 & 36.905 & +2.363 & 49.213 & 42.872 & +6.341 \\ 
CC+RSD+DESI DR2 & 51.284 & 53.027 & $-1.743$ & 62.736 & 59.898 & +2.838 \\ 
CC+RSD+DESI DR2+Union3 & 77.977 & 84.349 & $-6.372$ & 90.746 & 92.011 & $-1.264$ \\ 
\hline
\end{tabular}

\caption{AIC and BIC comparison for the VCDM and $\Lambda$CDM models. Here $\Delta \mathrm{AIC}=\mathrm{AIC}_{\mathrm{VCDM}}-\mathrm{AIC}_{\Lambda\mathrm{CDM}}$ and $\Delta \mathrm{BIC}=\mathrm{BIC}_{\mathrm{VCDM}}-\mathrm{BIC}_{\Lambda\mathrm{CDM}}$.}
\label{tab:aicbic}
\end{table}

\section{Conclusion}\label{conclusion}
In this study, we have investigated the cosmological consequences of a minimally modified gravity theory designed to reproduce VCDM-like behavior without invoking exotic matter components~\cite{DeFelice:2020prd, DeFelice:2020onz, DeFelice:2021xps, Ganz:2022zgs}. These theories belong to the Type-II class of MMG, characterized by the absence of an Einstein frame. Such frameworks introduce minimal deviations from General Relativity while preserving a stable and covariant formulation, making them promising candidates for explaining late-time cosmic acceleration. By incorporating an effective potential with specific functional dependence, the theory naturally gives rise to a transient feature in the Hubble expansion rate, providing a flexible mechanism to capture possible deviations from the standard cosmological model.

To test the observational viability of the model, we performed a comprehensive analysis using a combination of CC data, DESI DR2 BAO measurements, RSD growth data, and the Union3 Type Ia supernova sample. The model parameters were constrained using a Bayesian MCMC framework, ensuring a statistically robust estimation of confidence intervals. In addition to background probes, we incorporated the growth of cosmic structures through the observable \(f\sigma_8(z)\), enabling a direct comparison with large-scale structure data and allowing us to constrain the parameter \(S_8\).

Our results indicate that the VCDM model successfully reproduces the observed expansion and growth histories of the Universe. The Hubble parameter evolution exhibits a smooth transition feature near \(z \simeq 0.3\), associated with the model parameter \(A_2\), consistent with the vacuum metamorphosis mechanism. The inclusion of DESI DR2 and Union3 datasets refines this transition and improves overall agreement with observational data. Since the $A_2$ values are smaller for the final combination, the Hubble transition feature shifts toward a later epoch, indicating a delayed onset of the vacuum metamorphosis effect in the VCDM framework. The deceleration parameter \(q(z)\) and effective equation of state \(w_{\text{eff}}(z)\) show consistent behavior with a transition redshift around \(z_t \sim 0.7\), in excellent agreement with the standard \(\Lambda\)CDM model. The derived present-day values, such as \(q_0 \approx -0.56\) and \(w_{0} \approx -0.7\), confirm a late-time acceleration phase consistent with current constraints.

The growth analysis demonstrates that the VCDM framework reproduces the observed evolution of \(f\sigma_8(z)\) with high accuracy, yielding a present-day value of \(S_8 \simeq 0.80\), compatible with weak-lensing surveys and slightly lower than the Planck estimate, thus alleviating the well-known \(S_8\) tension. Moreover, a direct statistical comparison with \(\Lambda\)CDM using \(\chi^2\), AIC, and BIC criteria shows that the VCDM model achieves comparable or improved fits to the data, particularly when all probes (CC+RSD+DESI DR2+Union3) are considered jointly. In fact, the model outperforms \(\Lambda\)CDM in the statistical comparisons for the final combination. This highlights the model's ability to explain both geometric and growth observables with minimal theoretical modification.

Overall, this analysis demonstrates that Type-II minimally modified gravity theories exhibiting VCDM-like dynamics offer a viable and self-consistent extension of standard cosmology. They provide a natural mechanism for late-time acceleration, while maintaining compatibility with high-precision cosmological data. Future work may explore perturbative stability, non-linear structure formation, and potential observational discriminants in the upcoming DESI and Euclid datasets, which will further test the predictive power of such models.

\section*{Data availability} No new data was generated or analyzed to support this research.

\section*{Acknowledgement} SSM acknowledges the Council of Scientific and Industrial Research (CSIR), Govt. of India for awarding Junior Research fellowship (E-Certificate No.: JUN21C05815). SKB acknowledges the Council of Scientific and Industrial Research (CSIR), Govt. of India for awarding Junior Research fellowship (E-Certificate No.: 22D/23J06347). PKS acknowledges Anusandhan National Research Foundation (ANRF), Department of Science and Technology (DST), Government of India for financial support to carry out Research project No.: CRG/2022/001847 and IUCAA, Pune, India for providing support through the visiting Associateship program. We are very grateful to the honorable referees and the editor for the illuminating suggestions that have significantly improved our work in terms of research quality and presentation.

\bibliography{main}

\end{document}